\documentstyle[12pt]{article}

\catcode`\@=11
\long\def\@makefntext#1{
\protect\noindent \hbox to 3.2pt {\hskip-.9pt

$^{{\ninerm\@thefnmark}}$\hfil}#1\hfill}                

\def\@makefnmark{\hbox to 0pt{$^{\@thefnmark}$\hss}}  

\def\ps@myheadings{\let\@mkboth\@gobbletwo
\def\@oddhead{\hbox{}
\rightmark\hfil\ninerm\thepage}
\def\@oddfoot{}\def\@evenhead{\ninerm\thepage\hfil
\leftmark\hbox{}}\def\@evenfoot{}
\def\sectionmark##1{}\def\subsectionmark##1{}}

\renewcommand{\thefootnote}{\fnsymbol{footnote}}

\newcounter{sectionc}\newcounter{subsectionc}\newcounter{subsubsectionc}

\renewcommand{\section}[1] {\vspace*{0.6cm}\addtocounter{sectionc}{1}
\setcounter{subsectionc}{0}\setcounter{subsubsectionc}{0}\noindent
        {\normalsize\bf\thesectionc. #1}\par\vspace*{0.4cm}}

\renewcommand{\subsection}[1] {\vspace*{0.6cm}\addtocounter{subsectionc}{1}
        \setcounter{subsubsectionc}{0}\noindent
        {\normalsize\it\thesectionc.\thesubsectionc. #1}\par\vspace*{0.4cm}}

\renewcommand{\subsubsection}[1]
    {\vspace*{0.6cm}\addtocounter{subsubsectionc}{1}
     \noindent {\normalsize\rm\thesectionc.\thesubsectionc.\thesubsubsectionc.
        #1}\par\vspace*{0.4cm}}

\newcounter{appendixc}

\newcounter{subappendixc}[appendixc]

\newcounter{subsubappendixc}[subappendixc]

\renewcommand{\appendix}[1] {\vspace*{0.6cm}
        \refstepcounter{appendixc}
        \setcounter{figure}{0}
        \setcounter{table}{0}
        \setcounter{equation}{0}
        \renewcommand{\thefigure}{\Alph{appendixc}.\arabic{figure}}
        \renewcommand{\thetable}{\Alph{appendixc}.\arabic{table}}
        \renewcommand{\theappendixc}{\Alph{appendixc}}
        \renewcommand{\theequation}{\Alph{appendixc}.\arabic{equation}}
        \noindent{\bf Appendix \theappendixc #1}\par\vspace*{0.4cm}}

\def\abstracts#1{{
        \centering{\begin{minipage}{12.2truecm}\footnotesize\baselineskip=12pt\noindent
        \parindent=0pt #1
        \end{minipage}}\par}}

\renewenvironment{thebibliography}[1]
       {\begin{list}{\arabic{enumi}.}
        {\usecounter{enumi}\setlength{\parsep}{0pt}
\setlength{\leftmargin 1.25cm}{\rightmargin 0pt}
         \setlength{\itemsep}{0pt} \settowidth
        {\labelwidth}{#1.}\sloppy}}{\end{list}}

\newcounter{itemlistc}
\newcounter{romanlistc}
\newcounter{alphlistc}
\newcounter{arabiclistc}

\newcommand{\fcaption}[1]{
        \refstepcounter{figure}
        \setbox\@tempboxa = \hbox{\footnotesize Fig.~\thefigure. #1}
        \ifdim \wd\@tempboxa > 6in
           {\begin{center}
        \parbox{6in}{\footnotesize\baselineskip=12pt Fig.~\thefigure. #1}
            \end{center}}
        \else
             {\begin{center}
             {\footnotesize Fig.~\thefigure. #1}
              \end{center}}
        \fi}

\newcommand{\tcaption}[1]{
        \refstepcounter{table}
        \setbox\@tempboxa = \hbox{\footnotesize Table~\thetable. #1}
        \ifdim \wd\@tempboxa > 6in
           {\begin{center}
        \parbox{6in}{\footnotesize\baselineskip=12pt Table~\thetable. #1}
            \end{center}}
        \else
             {\begin{center}
             {\footnotesize Table~\thetable. #1}
              \end{center}}
        \fi}

\def\@citex[#1]#2{\if@filesw\immediate\write\@auxout
        {\string\citation{#2}}\fi
\def\@citea{}\@cite{\@for\@citeb:=#2\do
        {\@citea\def\@citea{,}\@ifundefined
        {b@\@citeb}{{\bf ?}\@warning
        {Citation `\@citeb' on page \thepage \space undefined}}
        {\csname b@\@citeb\endcsname}}}{#1}}

\newif\if@cghi
\def\cite{\@cghitrue\@ifnextchar [{\@tempswatrue
        \@citex}{\@tempswafalse\@citex[]}}
\def\citelow{\@cghifalse\@ifnextchar [{\@tempswatrue
        \@citex}{\@tempswafalse\@citex[]}}
\def\@cite#1#2{{$\null^{#1}$\if@tempswa\typeout
        {IJCGA warning: optional citation argument
        ignored: `#2'} \fi}}

 1
 1
 1

\font\ninerm=cmr9

\begin{document}


\begin{center}
{\normalsize\bf SINGLE-ELECTRON CHARGING \linebreak[4] OF SELF-ASSEMBLED InAs QUANTUM DOTS}
\end{center}
\vspace*{0.3cm}
\centerline{\footnotesize M. FRICKE, A. LORKE, M. HASLINGER AND J. P. KOTTHAUS}
\baselineskip=13pt
\centerline{\footnotesize\it Sektion Physik, LMU M\"unchen, Geschwister-Scholl-Platz 1, 80539 M\"unchen, Germany}
\vspace*{0.3cm}
\centerline{\footnotesize G. MEDEIROS-RIBEIRO and P. M. PETROFF}
\baselineskip=13pt
\centerline{\footnotesize\it Materials Department and QUEST, University of California, Santa Barbara, CA 93106, USA}

\vspace*{0.9cm}

\abstracts{The many-particle ground state and excitations of self-assembled InAs quantum dots are investigated using far-infrared (FIR) and capacitance spectroscopy. The contributions of quantization and charging energies to the $n$--electron energy spectrum are determined. For $n = 1$, 2, we find good agreement with a parabolic model; for higher filling, the Coulomb interaction reveals itself not only in the capacitance, but also in the FIR excitations.}

\vspace*{0.6cm}
\normalsize\baselineskip=15pt
\setcounter{footnote}{0}
\renewcommand{\thefootnote}{\alph{footnote}}

The mechanisms that lead to the formation of three-dimensional islands in the epitaxial growth of strongly lattice-mismatched materials have been known for several decades.\cite{Stranski} However, only in recent years it has been realized that this growth mode is suitable for the fabrication of well defined, nm-size quantum dots in the 10 nm range.\cite{VierSAD,Drexler,Medeiros,Leonard} This size is of particular interest, since the three-dimensional carrier confinement that is present in such dots leads to energy quantization that is large enough to make use of the associated $\delta$-like density of states even at elevated temperatures. Furthermore, quantization energies and Coulomb contributions to the energetic structure of the carrier system in such small dots are of comparable size.

Here, we investigate self-assembled InAs quantum dots by capacitance and far-infrared (FIR) spectroscopy. Whereas the former experimental technique probes the ground-state of the many-electron system in the dots, the latter is sensitive to its excitations. The combination of both techniques therefore allows for a detailed investigation of the different contributions to the electronic structure of the dots.

The samples are grown by molecular beam epitaxy on semi-insulating GaAs. The InAs dots are embedded into a metal-insulator-semiconductor field-effect-transistor (MISFET) structure, which allows for an {\it in situ} tuning of the electron number per dot by  application of a suitable gate voltage. The MISFET structure consists of a heavily doped back contact, 175 nm below the surface, 25 nm GaAs spacer layer, the InAs self-assembled dots with the associated wetting layer, 30 nm GaAs spacer, 116 nm Al$_{0.75}$Ga$_{0.25}$As digital alloy barrier layer, followed by a 4 nm GaAs cap. Details of the growth procedure used for the formation of the InAs dots can be found, e.g., in Refs.\ \cite{Leonard,Medeiros}. We estimate the dots' diameter and height to be 20 nm and 7 nm, respectively.  

Small samples are cleaved from the wafer, wedged, and provided with Ohmic contacts and a semi-transparent gate. They are then mounted in a liquid-He cryostat, equipped with a superconducting solenoid, capable of producing fields of up to $B = 15$ T, perpendicular ($\perp$) as well as parallel ($\parallel$) to the plane of the dots.

The FIR transmission of the sample was recorded using a rapid-scan Fourier transform spectrometer, the capacitance was measured {\it in situ} by standard lock-in technique.

\begin{figure}
\vspace{5.5 cm}
\fcaption{Capacitance--voltage traces at $B = 0$ and $B_{\perp} =12$ T. Numbers indicate electrons per dot. Curves are offset for clarity.}
\label{FigKap}
\end{figure}

Figure 1 displays typical capacitance traces of our samples at $B = 0$ and $B_\perp = 12$~T. At very low gate bias, $V_g \leq -1$ V, the signal is given by the geometric capacitance between the top gate and the doped GaAs back contact. At  $V_g = -0.9$ V a sharp increase of the capacitance indicates the charging of the lowest electron state in the dots.\cite{Drexler} Even though this state (the {\it s}-state) is doubly spin degenerate, the second electron is loaded at a somewhat higher gate voltage of $-0.7$ V, because the 2-electron ground state is affected by the repulsive electron--electron interaction, which leads to the well-known Coulomb blockade effect.\cite{CoulombB} 

Taking into account the image charge induced in the metallic back gate, the difference between 1-- and 2--electron ground state, $E_{1,2}$, can be related to the gate voltage difference $\Delta V_g$ between charging peaks by

\begin{equation}
E_{1,2}=e\frac{t_b}{t_{tot}}\Delta V_g + \frac{e^2}{8\pi \varepsilon \varepsilon_0 t_b}
\label{eq2}\end{equation}

Here, $t_b$ ist the distance between the back contact and the dots and $t_{tot}$ is the distance between the back contact and the gate. For the present sample, with $t_b = 25$ nm and $t_b / t_{tot} = 1/7$, we find $E_{1,2} = 23.3$ meV. 

In our oblate, nearly circular dots, the next higher ({\it p}--) state is expected to be fourfold degenerate at $B=0$, so that the Coulomb charging energy is expected to result in four, roughly equidistant charging peaks. For the present sample, these have merged into one broad resonance (in small-scale samples with less inhomogeneous broadening, however, the individual peaks can be resolved \cite{Miller}). When a perpendicular magnetic field is applied, the Zeeman term leads to a splitting of the {\it p}--state, with two electron levels moving down and two electrons moving up with increasing field.\cite{Hansen} This can  be observed in the upper trace in Fig.\ 1. The combined energetic shift is directly given by the cyclotron energy $\hbar \omega_c = \hbar eB/m^*$,\cite{Hansen,MerktundCo} which allows us to extract an effective mass of $m^* = 0.066 \pm 0.015$ $m_e$. This value is considerably higher than the conduction band edge mass of InAs, which is partly due to the high nonparabolicity of this material and partly to the fact that a large fraction of the electron wave function leaks into the GaAs barrier layers. 

\begin{figure}
\vspace{5.5 cm}
\fcaption{(a) Normalized ground state energy difference $E_{1,2}$ as a function of magnetic field. (b) Contribution to $E_{1,2}$ from magnetic-field-induced compression. }
\label{FigSq}
\end{figure}

As seen in Fig.\ 2, $ E_{1,2}$ is almost independent of magnetic field. Careful evaluation of the peak structure, however, shows that $E_{1,2}$ increases by $\approx 2$\% when a magnetic field of 12 T is applied perpendicular to the plane of the dots (solid data points in Fig.\ \ref{FigSq}(a)). Apart from spin-splitting, this shift can be attributed to an increase of the Coulomb energy caused by a magnetic-field-induced compression of the wave function. Following an argument by Merkt {\it et al.},\cite{MerktundCo} we calculate the Coulomb interaction in parabolic approximation, $E_{1,2}^{e-e} =e^2/(4\pi \varepsilon\varepsilon_0 \ell)$, with a characteristic length of the ground state wave function, $\ell=\sqrt{\hbar/(m^* \omega)}$, which is magnetic field dependent through $\omega = \sqrt{\omega_0^2 + \omega_c^2 / 4}$ (solid line).


The contributions from spin and magnetic compression can be distinguished by their dependence on the direction of $B$: Because of the large confinement in the growth direction, only the spin splitting contributes to $E_{1,2}$ for parallel magnetic fields and the associated shift (open symbols) is only about half of that for perpendicular field. As shown in Fig.\ 2(b), the difference between $E_{1,2}(B_\parallel)$ and $E_{1,2}(B_\perp)$ can be well explained by a  magnetic-field-induced compression of the ground state wave function, using the above expressions for $E_{1,2}^{e-e}$. From the splitting in a parallel field, $E_{1,2}(B_\parallel)$, we extract an effective $g$--factor of $g_{dot} \approx 0.4$.

\begin{figure}
\vspace{5.5 cm}
\label{FigFIRpos}
\fcaption{Far-infrared resonance positions as a function of magnetic field.}
\end{figure}

Figure 3 gives an overview of the measured FIR resonance positions as a function of magnetic field for different electron occupation numbers. For $n_e = 1$, 2 (Fig.\ 3(a)) we find the characteristic two-mode spectrum of a parabolically confined electron system (solid lines) with a confinement energy of $\hbar \omega_0 = 49$ meV. 

This situation changes drastically, when the p-- state starts filling and transitions between higher lying states become possible.\cite{Fricke} As seen in Fig.\ 3(b), then the upper mode, $\omega_+$, splits up into three resonances. This behavior of the $\omega_+$--mode can be observed for $n_e = 3$, 4 and 5 and shows the importance of electron--electron interactions when the $p$--shell is partially filled. For  $n_e = 6$ we again observe just two resonances, however, with a dispersion that strongly deviates from that of a parabolic dot. 

In comparing the results from capacitance and FIR spectroscopyy, we find good agreement between the characteristic length derived from $\ell = e^2/(4\pi \varepsilon\varepsilon_0 E_{1,2}) = 4.9$ nm and $\ell=\sqrt{\hbar/(m^* \omega_0)} = 4.4$ nm. This nicely demonstrates the applicability of the models used and shows the compatibility of both experimental techniques


We gratefully acknowledge financial support through QUEST, a NSF Science and Technology Center and through BMBF grant \#1BM623.

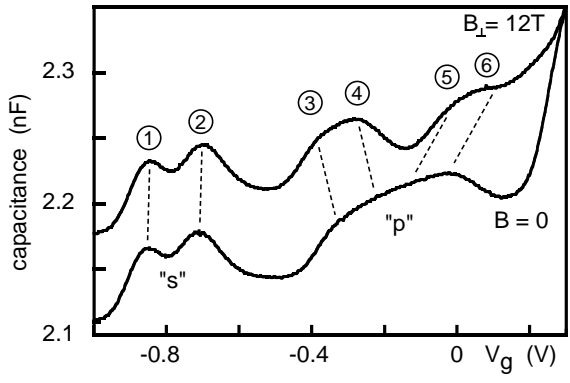
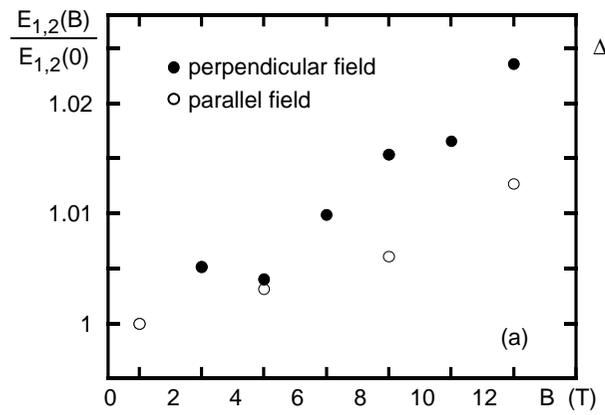
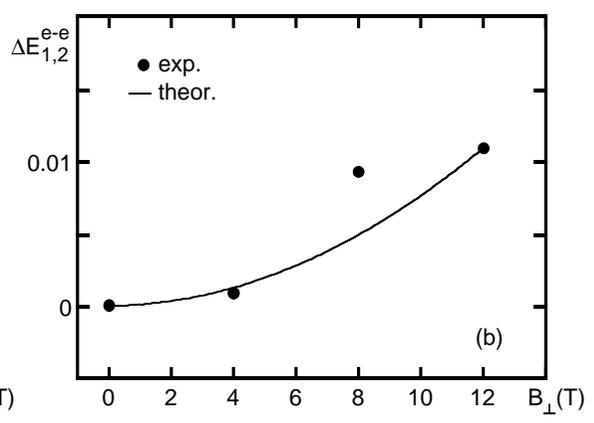
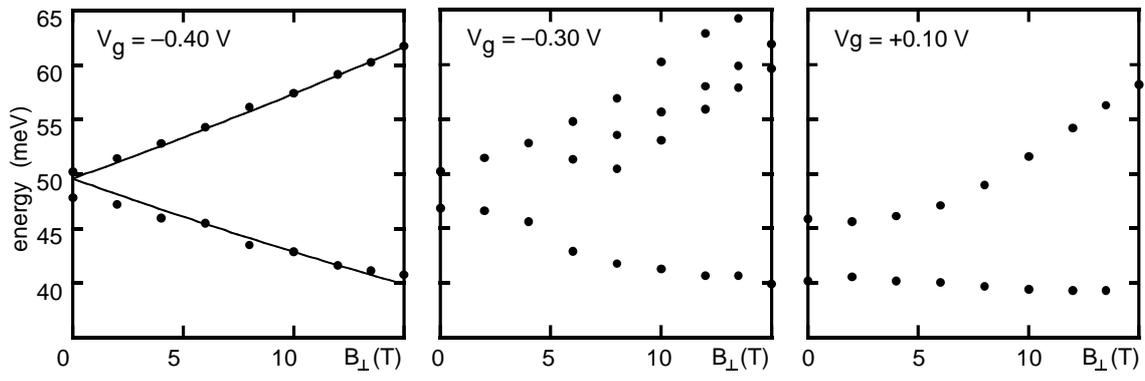
\end{document}